\documentstyle[preprint,aps]{revtex}
\begin{document}
\draft
\preprint{\parbox{4cm}{\flushright CLNS 95/1377\\ MZ-TH/95-31}}
\title{Sum Rules for Radiative and Strong Decays of Heavy Mesons}
\author{Chi-Keung Chow}
\address{Newman Laboratory of Nuclear Studies, Cornell University, Ithaca,
NY 14853.}
\author{Dan Pirjol\cite{0}}
\address{Johannes Gutenberg-Universit\"at, Institut f\"ur Physik (THEP),\\
Staudingerweg 7, D-55099 Mainz, Germany.}
\date{\today}
\maketitle
\begin{abstract}
We derive two model-independent sum rules relating the transition matrix
elements for
radiative and strong decays of excited heavy mesons to properties of
the lowest-lying heavy mesons. The sum rule for the radiative decays is an
analog of the
Cabibbo-Radicati sum rule and expresses the sum of the radiative widths in
terms of the isovector charge radius of the ground state heavy meson.
Using model-dependent estimates and heavy hadron chiral perturbation
theory calculations, we show that this sum rule is close to saturation
with states of excitation energies less than 1 GeV.
An analog of the Adler-Weisberger sum rule gives an useful sum rule for
the pionic widths of heavy excited mesons, which is used to set a
model-independent upper bound on the coupling of the P-wave heavy mesons.
\end{abstract}
\narrowtext
  We present in this paper two model-independent sum rules for the photon
and pion couplings to the light constituents in a heavy meson. They relate
properties of the ground state such as charge radius, magnetic moment
and axial coupling to matrix elements which govern the radiative and
strong transitions between excited heavy meson states and the ground
state. The latter parameters can be in principle extracted from experiment
and in fact, a few of them have been already determined in this way.
The sum rules themselves are not new, they are direct analogs of the
Cabibbo-Radicati (CR) \cite{CR} and respectively Adler-Weisberger (AW)
\cite{A,We} sum rules, familiar from
current algebra. However, in this new context they turn out to be considerably
more predictive than in their original application, due to additional
constraints on the quantum numbers of the available final states. Moreover,
when considered in the large-$N_c$ limit, the form of the sum rules simplifies
further due to the suppression of the continuum contribution. The
CR sum rule reduces in this limit to a constituent-quark sum
rule familiar from nonrelativistic quantum mechanics, connecting the
charge radius of the ground state to a sum over electric dipole matrix
elements between excited states and the ground state \cite{BS}.

These sum rules are of interest from a phenomenological point of view,
as they can be used to place constraints on the photon and pion couplings
of low-lying heavy mesons to excited ones.
As a sample application, we derive model-independent upper bounds on the
pionic decay widths of the charmed p-wave heavy mesons with the quantum
numbers of the light degrees of freedom $s_\ell^{\pi_\ell}=1/2^+$.

The Cabibbo-Radicati sum rule can be derived (we follow here the derivation
given in \cite{Beg,Kaw,W}) by considering the forward
scattering amplitude of isovector photons of energy $\omega$ and helicity
$\lambda$ on a target $B$
\begin{eqnarray}
f^{ab}(\omega,\lambda)
= \frac{i}{4\pi}\int\mbox{d}x\,e^{-iq\cdot x}
\langle B|\mbox{T}(J^a\cdot e_\lambda)(x)(J^b\cdot e^*_\lambda)(0)|B
\rangle\,,
\end{eqnarray}
with $q^0=\omega$, $q^2=0$ and $J^a_\mu=e\bar q\gamma_\mu
t^aq$ ($t^a=\tau^a/2$). The states $|B\rangle$ are normalized
noncovariantly to 1.
The scattering amplitude $f$ can be written in terms
of four invariants $f^{(\pm)}_\pm(\omega)$ as ($J_z$ is the $z$-projection
of the target spin)
\begin{eqnarray}
f^{ab}(\omega,\lambda)
&&= \langle B|\frac12\{t^a,t^b\}\left(f^{(+)}_+(\omega)
+ f^{(+)}_-(\omega)\lambda\omega J_z\right) |B\rangle\nonumber\\
+&& \langle B|\frac12[t^a,t^b]\left(f^{(-)}_+
(\omega)\lambda J_z + \omega f^{(-)}_-(\omega)\right)|B\rangle\,.
\end{eqnarray}
The assumption of an unsubtracted dispersion relation for $f^{(-)}_-
(\omega)$
\begin{eqnarray}
f^{(-)}_-(0) = \frac{2}{\pi}\int_0^\infty\frac{\mbox{d}\omega'}{\omega'}
\mbox{Im }f^{(-)}_-(\omega')
\end{eqnarray}
in combination with the low-energy theorem \cite{Beg,Kaw,W}
$\lim_{\omega\to 0}f^{(-)}_-(\omega) = \frac{1}{\pi}(-R_V^2/6 + \mu_V^2/2)$
(for a target of isospin $\frac12$ and spin $\frac12$)
leads to the final form of the CR sum rule
\begin{eqnarray}\label{CR}
\frac{R_V^2}{6} - \frac{\mu_V^2}{2} = \frac{1}{\pi^2 e^2}
\int_0^\infty\frac{\mbox{d}\omega'}{\omega'}
\left(2\sigma_{1/2}(\omega') - \sigma_{3/2}(\omega')\right)\,.
\end{eqnarray}
On the left-hand side $R_V$ and $\mu_V$ are the isovector charge radius and
magnetic moment of the target. On the right-hand side the optical theorem
has been used to express Im$\,f^{(-)}_-(\omega)$ in terms of the
cross-sections for inclusive photoproduction by an isovector photon with
$I_z=0$ of final states with isospin 1/2 and respectively 3/2.

We will take as target a pseudoscalar heavy meson with quark
content $\bar Qu$ $(I,I_z=\frac12,+\frac12)$, denoted generically
as $|B_i\rangle$ ($i=u,d$).
There are a number of specific points which have to be addressed in
connection with this choice.
First, the sum rule (\ref{CR}) has been derived under the assumption that
the target is a spin-1/2 particle, whereas the $|B_i\rangle$ meson has spin
zero.
However, in the heavy mass limit $m_Q\to\infty$ the dynamics of the heavy
quark decouples from that of the light constituents. As a
result, the target $|B_i\rangle = \frac{1}{\sqrt{2}}(|Q^\uparrow
q_i^\downarrow
\rangle - |Q^\downarrow q_i^\uparrow\rangle)$ can be effectively considered
as a coherent superposition of polarized spin-1/2 particles.

   Second, for such a target the isovector and isoscalar electromagnetic
parameters are related in the SU(3)-symmetric limit. The reason for this
is that
heavy mesons containing only one light quark transform according to the {\bf 3}
representation of SU(3). The electromagnetic current transforms as an octet
and there is only one way of combining {\bf 3}, {\bf 8} and ${\bf\bar 3}$ to
a singlet \cite{CG,Am,Yan}. Previous applications
of the CR sum rule used a proton or a pion as target, which belong to
SU(3) octets. Therefore their isovector and isoscalar electromagnetic
form-factors remain unrelated even in the SU(3) limit. The assumption of
SU(3) symmetry simplifies very much the sum rule, and the main part of the
discussion below will be restricted to this case.
However, the sum rule can be modified to include SU(3)-breaking effects and
we will return to this point later.

 From the above observation it follows that the elastic e.m. form-factor of
a B meson can
be written in terms of just one function $F(q^2)$ as
$\langle B_i|J_\mu|B_j\rangle = eF(q^2)v_\mu\, {\cal Q}_{ij}$
with $J_\mu=e\bar q\gamma_\mu {\cal Q}q$ and ${\cal Q}=$ diag($\frac23,-\frac13
$) is the light quark charge matrix. Current conservation gives $F(0)=1$.
The charge radius appearing on the l.h.s. of (\ref{CR}) is defined as
$R_V^2/6 = dF(q^2)/dq^2|_{q^2=0}$.

  The isovector magnetic moment $\mu_V$ of the light constituents in
(\ref{CR})
is related to the parameter $\beta$ introduced in \cite{Am}\footnote{The
extra (--) sign is due to the fact that we are considering heavy
mesons with one heavy antiquark.}
which describes the radiative decay $B^*\to B\gamma$
$\langle B(v)|\bar q\gamma_\mu q|B^*(v,\epsilon)\rangle =
-i\beta\varepsilon_{\mu\nu\rho\sigma}v^\nu k^\rho \epsilon^\sigma$
($k$ is the photon momentum) as $\mu_V=\beta/2$. The corresponding decay
rate is equal to $\Gamma = \frac13\alpha Q^2\beta^2|\vec k\,|^3$, with
$Q$ the heavy quark charge in units of $e$.

  Finally, another distinctive feature of our problem is the absence of
resonant final states with isospin 3/2. Therefore, $\sigma_{3/2}(\omega)$
in (\ref{CR}) only receives contributions from continuum states.
There is one limit in which these contributions are completely suppressed,
and this is the large-$N_c$ limit. In this case the Cabibbo-Radicati
sum rule is saturated with the resonances alone, and the corresponding
photoproduction cross-sections can be expressed in terms of the decay
widths for the inverse process $B^{exc}\to B\gamma$. One obtains in this
way the particularly simple result
\begin{eqnarray}\label{largeNc}
\frac{R_V^2}{6} - \frac{\beta^2}{8} = \frac{1}{8\alpha Q^2}
\sum_{exc}(2J+1)\frac{\Gamma(B^{exc}\to B\gamma)}{|\vec k\,|^3}\,.
\end{eqnarray}
The summation runs over all excited states of the $|B\rangle$ meson and $J$
is the spin of each state. The form of the sum rule can be simplified
by deleting the second term on the l.h.s. and extending the
summation over the $B^*$ state as well.

  It is interesting that a very similar sum rule can be obtained in the
nonrelativistic constituent quark model (NRCQM) \cite{NRCQM}
(and also in atomic physics \cite{BS}) by writing
\begin{eqnarray}\label{R2}
\langle B|\vec x\cdot\vec x|B\rangle = \sum_n^{(\ell=1)}
|\langle B|\vec x|n\rangle|^2\,.
\end{eqnarray}
The radiative decay rate in the electric dipole approximation is given by
$\Gamma_{E1}=\frac43\alpha Q^2|\vec k\,|^3|\langle B|\vec x|n\rangle|^2$, which
yields upon insertion into (\ref{R2}) a sum rule for the heavy meson charge
radius
\begin{eqnarray}\label{NRQM}
\frac{R^2}{6} = \frac{1}{8\alpha Q^2}\sum_{exc}^{(\ell=1)}(2J+1)
\frac{\Gamma_{E1}(B^{exc}\to B\gamma)}{|\vec k\,|^3}\,.
\end{eqnarray}
Although very similar to (\ref{largeNc}), the summation over excited states
extends in (\ref{NRQM}) only over P-wave states, which are connected to the
ground state by a E1 transition, whereas in the exact sum rule
(\ref{largeNc}) {\em all} excited states contribute.

 When the large-$N_c$ limit is relaxed, the integral on the right-hand side
of the sum rule (\ref{CR}) will receive, in addition to the contributions
from the excited heavy mesons (\ref{largeNc}), also contributions from
continuum states like ($B\pi$), ($B^*\pi$), etc.
These can be calculated reliably in heavy hadron chiral perturbation theory
\cite{HHCPT1,HHCPT2,HHCPT3}, as long as the pion momentum is smaller than
the chiral
symmetry breaking scale $\Lambda_\chi\simeq 1$ GeV. By keeping only the
B and B$^*$ mesons in intermediate states we obtain
\begin{eqnarray}\label{sig1}
\sigma_{1/2}^{(\pi B)}(\omega) &=& 2\sigma_{3/2}^{(\pi B)}(\omega) =
\frac{\pi}{18}\left(\frac{eg\beta}{2\pi f_\pi}\right)^2
\frac{|\vec p\,|}{\omega}(\omega^2-m_\pi^2)\,,\\
\sigma_{1/2}^{(\pi B^*)}(\omega) &=& \frac{2\pi}{3}\left(\frac{eg}{2\pi f_\pi}
\right)^2\frac{|\vec p\,|}{\omega}\left\{ 1 + \frac{m_\pi^2}{\omega^2} +
\frac{m_\pi^2}{2\omega|\vec p\,|}\log\frac{\omega-|\vec p\,|}
{\omega+|\vec p\,|}\right.\\
 &-&\,\left. \frac14\beta\omega\left(1 +
\frac{m_\pi^2}{2\omega|\vec p\,|}\log\frac{\omega-|\vec p\,|}
{\omega+|\vec p\,|}\right) +
\frac{1}{24}\beta^2(\omega^2-m_\pi^2)\right\}\,,\nonumber\\
\sigma_{3/2}^{(\pi B^*)}(\omega) &=& \frac{\pi}{3}\left(\frac{eg}{2\pi
f_\pi}\right)^2\frac{|\vec p\,|}{\omega}\left\{1 + \frac{m_\pi^2}
{\omega^2} + \frac{m_\pi^2}{2\omega|\vec p\,|}\log\frac{\omega-|\vec p\,|}
{\omega+|\vec p\,|}\label{sig3}\right.\\
 &+&\,\left. \frac12\beta\omega\left(1 +
\frac{m_\pi^2}{2\omega|\vec p\,|}\log\frac{\omega-|\vec p\,|}
{\omega+|\vec p\,|}\right) +
\frac16\beta^2(\omega^2-m_\pi^2)\right\}\,,\nonumber
\end{eqnarray}
where $|\vec p\,|=\sqrt{\omega^2-m_\pi^2}$ is the pion momentum.
The static approximation for the heavy meson has been used in deriving
these expressions $m_\pi/m_B\simeq 0$.

In these formulas $g$ is the BB$^*\pi$ coupling in the heavy mass limit
\cite{HHCPT1,HHCPT2,HHCPT3}. Experimental data on branching
ratios of D$^*$ decays \cite{CG,Am,Yan} give $0.09 \leq g^2 \leq 0.5$ (with
90\% confidence limits). We will adopt for our estimates the upper
limit $g^2=0.5$ (note though that QCD sum rule computations suggest
significantly lower values $g^2\simeq 0.1$ \cite{Col,GY,Bel,DN}).
As already mentioned, $\beta$ describes the $B^*\to B\gamma$ decay in the
heavy mass limit.
It has been determined simultaneously with $g$ in \cite{CG,Am,Yan} such
that their values are correlated: larger values for $g$ favor larger values
for $\beta$.
The limits quoted are $2\leq\beta\leq 6$ GeV$^{-1}$ \cite{CG,Am,Yan}.
Recent QCD sum rule and model calculations \cite{Aliev,Col,DN} give on
the other hand smaller values, around $\beta=2$--$3$
GeV$^{-1}$ (\cite{DN} find $\beta\simeq 1$ GeV$^{-1}$).
We will use in our estimates below $\beta=3$--$4$ GeV$^{-1}$.

Previous experience with the Cabibbo-Radicati sum rule \cite{CR1,CR2,CR3}
suggests that the saturation region includes states with an excitation
energy of the order of a few GeV. Inserting (\ref{sig1}-\ref{sig3}) into
(\ref{CR}) with an upper cutoff of 1 GeV gives for the continuum
contribution on the r.h.s. of the sum rule (\ref{CR})
\begin{eqnarray}
I_{cont}^{(\pi B)}+ I_{cont}^{(\pi B^*)} + \cdots
&=& \left(\frac{g}{2\pi f_\pi}\right)^2 0.036\beta^2 +
\left(\frac{g}{2\pi f_\pi}\right)^2(1.519 - 0.321\beta)
 + \cdots\nonumber\\
& =& 0.639 (0.589)\,(\mbox{GeV}^{-2})+ \cdots\,.
\end{eqnarray}
The ellipsis stand for other continuum contributions, which are expected
to be less important as they have less phase space available. The two
numbers correspond to $\beta = 3(4)$ GeV$^{-1}$. The pion decay constant
is $f_\pi=0.132$ GeV.

  We are now in a position to discuss the numerical values of the two
sides of the sum rule (\ref{CR}). It can be written as
\widetext
\begin{eqnarray}\label{numbers}
2.161 &=&
0.321 (0.846) + (0.405\pm 0.067) + (0.811\pm 0.135) + 0.639 (0.589) +
\cdots\,.(\mbox{GeV}^{-2})\nonumber\\
\frac{R_V^2}{6}\quad &=&\quad (B^*)\qquad + \qquad (P_{1/2})\qquad +
\qquad (P_{3/2}) \quad + \quad (\mbox{continuum})
\end{eqnarray}
\narrowtext

On the left-hand side, the isovector charge radius $R_V$ has been obtained
from a version of vector-meson dominance where the contributions of the two
lowest $I=1$ $(J^{PC}=1^{--})$ vector mesons are kept
$R_V^2/6=(1/m_\rho^2 + 1/m_{\rho'}^2)$.
This procedure gives
an elastic form-factor in agreement with the QCD counting rules at large
$q^2$ \cite{LS}.
On the right-hand side, the B$^*$ contribution has been computed by
including also the nonanalytic SU(3)-violating contributions obtained
in \cite{Am} $\mu_V^2 = \frac14(\beta-\frac{g^2m_K}{4\pi f_K^2}-
\frac{g^2m_\pi}{2\pi f_\pi^2})^2$. The contributions of the two lowest-lying
P-wave states with $s_\ell^{\pi_\ell}=\frac12^+,\frac32^+$ were computed
in the dipole approximation
using the wavefunctions of the ISGW model \cite{ISGW} with the updated
parameters of the ISGW2 model \cite{ISGW2}. In the absence of the spin-orbit
interaction, responsible for the splitting of these two multiplets, their
contributions to the sum rule are in the ratio $P_{1/2}:P_{3/2}=1:2$.
The errors shown correspond to the 30\% accuracy expected from the model
when predicting radiative decay matrix elements \cite{ISGW2}.

  One obtains in this way for the r.h.s.of the sum rule $2.176\pm 0.202$
$(2.651\pm 0.202)$ GeV$^{-2}$. The agreement with the l.h.s. is certainly
better than one could have expected from our qualitative estimates. This
strongly suggests that the
CR sum rule is very close to saturation with the first few excited states
and the continuum up to excitation energies of about 1 GeV.

 We would like to make a few additional comments about the different
contributions on the r.h.s. of (\ref{numbers}).
\begin{itemize}
\item The nonresonant contributions have all the same sign (positive).
The dominant contribution to the integral over $\omega$ comes from the
$(\pi B^*)$ two-body state, which can be formed in a $l=0$ partial wave.
Therefore this cross-section dominates in the low energy region, to which
the sum rule is the most sensitive. The threshold region
gives rise to a large log $(g/(2\pi f_\pi))^2\log(\Lambda_\chi/m_\pi)$,
which is consistent with the expected divergence of the isovector charge
radius in the chiral limit $m_\pi\to 0$ \cite{BZ}.

\item An important issue which needs to be addressed is that of the
contributions of higher excited states. We have estimated these
contributions with the help of the simpler NRCQM sum rule (\ref{NRQM}).
In general,
the approach to saturation depends on the specific potential adopted.
For an harmonic oscillator the sum rule is saturated with the first
P-wave states alone and for a particle in a spherical potential well these
states contribute over 99\% of the total. A similar behaviour is
expected to be true of any other confining potential. On the other hand,
in the hydrogen atom the lowest-lying P-wave states contribute 55\%
and the continuum states about 28\% of the total \cite{BS}.

\item The r.h.s. of (\ref{numbers}) is an increasing function of $\beta$,
with a minimum at $\beta=1.93$ (for $g^2=0.5$). Therefore requiring equality
of the two sides seems to favor
values for $\beta$ of the order of 3 GeV$^{-1}$. However, the estimate
of the P-wave contributions is still too crude to allow setting an useful
constraint on $\beta$. In case that these contributions turn out to have
been overestimated, the sum rule will be well satisfied with larger values
of $\beta$ ($~4$--$6$ GeV$^{-1}$).
\end{itemize}

  A discussion of the SU(3)-breaking effects is now in order. SU(3) was
needed to express the photoproduction cross-section $\sigma_{1/2}(\omega)$
in (\ref{CR}) in terms of the decay widths $\Gamma$ in (\ref{largeNc}).
When it is broken, the relation between these two is not anymore a simple
one: the former quantity only depends on the isovector properties of the
excited state, whereas the latter contains both isovector and isoscalar
contributions (unrelated to each other).

  However, let us assume that the individual
partial widths of all multipoles allowed in a given radiative decay
$\Gamma=\sum_i \Gamma_i$ will be measured experimentally. Then it will be
possible to eliminate the isoscalar
transition matrix element between the same partial widths for the two members
of the isospin doublet $\Gamma_i^{(u)}$, $\Gamma_i^{(d)}$.
One obtains in this way
\begin{eqnarray}\label{subtr}
\sigma_{1/2}(\omega) = \frac{\pi^2}{4}
\sum_{exc}(2J+1)\sum_i\frac{(\sqrt{\Gamma_i^{(u)}}-\sqrt{\Gamma_i^{(d)}})^2}
{|\vec k\,|^3}\delta(\omega-\delta M)\,.
\end{eqnarray}
The determinations of the square roots must be chosen which satisfy
$\sqrt{\Gamma_i^{(u)}}:\sqrt{\Gamma_i^{(d)}}$ = 2/3 : --1/3 in
the SU(3)-symmetric limit. With this choice (\ref{largeNc}) is
recovered in the SU(3) limit. This method requires a good theoretical
control over the signs of the matrix elements involved. Also from the
experimental point of view the measurement of the different partial widths
might appear to be very difficult. However, if successful, the subtraction in
(\ref{subtr}) will eliminate at the same time with the matrix element of
the isoscalar light quark e.m. current, also the contribution from the
heavy quark.
This will extend the domain of applicability of the sum rule to the charm
sector, for which the validity of the heavy mass expansion is less certain.

  Finally, we note that the methods of this paper can be used with little
modification to obtain a sum rule for the pion couplings between heavy
excited mesons and the ground state ones $B,B^*$. To derive it, consider
the amplitude for forward scattering of pions on a $B$ meson.
The assumption of an unsubtracted dispersion relation for the isospin-odd
part of this amplitude plus knowledge of its low-energy limit gives an
analog of the well-known Adler-Weisberger sum rule \cite{A,We}
\begin{eqnarray}\label{AW}
1-g^2 =
\frac{f_\pi^2}{\pi}\int_{m_\pi}^\infty\frac{\mbox{d}\nu}{\nu^2}\sqrt{\nu^2-
m_\pi^2}\left(\sigma(\pi^-B_u\to X) - \sigma(\pi^+B_u\to X)\right)\,.
\end{eqnarray}
On the l.h.s. $g$ is the BB$^*\pi$ coupling defined as above; the integral
on the r.h.s. runs over the inclusive cross-sections for $B_u\pi^\pm$
scattering with energy $\nu$. Just as in the case of the CR sum rule,
the heavy resonances will contribute only to $\sigma(\pi^-B_u\to X)$,
since the $\pi^+B_u$ state has isospin 3/2. Separating explicitly
the contribution of the resonances from the continuum $(I_{cont})$, the
sum rule (\ref{AW}) can be written as
\begin{eqnarray}
1 = 2\pi f_\pi^2 \sum_{res}(2J+1)\frac{\Gamma(B_d^{res}\to \pi^-B_u)}
{\nu^3} + I_{cont}\,.
\end{eqnarray}
Even without a detailed calculation one can see that $I_{cont}$ is positive,
because $B_u\pi^-$ has simply more available channels than $B_u\pi^+$.
This observation can be used to produce a model-independent constraint on the
couplings of the higher resonances:
\begin{eqnarray}\label{AWnumbers}
& & g^2 + |h|^2 + (0.07\pm 0.01) + \cdots < 1\\
& &\hspace{-0.2cm}(B^*)\,\,\, P_{1/2}\qquad\,\,\, P_{3/2}\nonumber
\end{eqnarray}
The contribution of the $s_\ell^{\pi_\ell}=3/2^+$ heavy mesons has been
obtained from existing data on their decay widths \cite{KKP,FL}.
On the other hand, the members of the $s_\ell^{\pi_\ell}=1/2^+$ multiplet
whose strong
couplings are parametrized by $h$ (defined as in \cite{F}) have not been
observed experimentally due to their large width.
The relation (\ref{AWnumbers}) gives an useful upper bound on $|h|$:
$|h|^2<0.5$, where we neglected the small contribution from the
other excited states and used $g^2=0.5$ as above. This agrees well with the
recent QCD sum rule calculation of \cite{CFN}, who find $|h|^2=0.39\pm 0.31$.
For the total pionic widths of the $0^+$ and $1^+$ charmed states our result
implies the
upper bounds $\Gamma(0^+)\leq 328$--$560$ MeV and $\Gamma(1^+)\leq
110$--$250$ MeV, corresponding to the mass values $m_{0^+}=m_{1^+} =
2.3$--$2.4$ GeV \cite{P12}.

\acknowledgments
C.K.C. thanks Peter Lepage, Mark Wise and Tung-Mow Yan for discussion.
His work is supported by the National Science Foundation.
D.P. thanks J\"urgen K\"orner and Karl Schilcher for useful discussions on
the subject of this paper. He acknowledges a grant from the Deutsche
Forschungsgemeinschaft (DPG).


\end{document}